\documentclass[
 reprint,
 amsmath,amssymb, aps, pra,superscriptaddress,longbibliography]{revtex4-1}
\usepackage{setspace}
\usepackage{emptypage}
\usepackage{makeidx}
\usepackage{graphicx, subcaption}
\graphicspath{ {./figures/} }
\usepackage[justification=raggedright]{caption}
\usepackage{enumitem}
\usepackage{wrapfig}
\usepackage{verbatim}
\usepackage[toc,page]{appendix}
\usepackage{commath}
\usepackage{tensor}
\usepackage{stackengine}
\usepackage{xcolor,soul}
\usepackage{mathrsfs}
\usepackage{dcolumn}
\usepackage{bm}
\usepackage[english]{babel}
\DeclareUnicodeCharacter{2009}{}
\usepackage[bookmarks=false,unicode=true,colorlinks=true,linkcolor=black,citecolor=black,urlcolor=black,breaklinks=true]{hyperref}

\begin{document}
\title{Dielectric-Boosted Sensitivity to Cylindrical Azimuthally Varying Transverse-Magnetic Resonant Modes in an Axion Haloscope}
\author{Aaron P. Quiskamp}
\email{aaron.quiskamp@research.uwa.edu.au}
\affiliation{ARC Centre of Excellence for Engineered Quantum Systems and ARC Centre of Excellence for Dark Matter Particle Physics, Department of Physics, University of Western Australia, 35 Stirling Highway, Crawley, WA 6009, Australia.}
\author{Ben T. McAllister}
\affiliation{ARC Centre of Excellence for Engineered Quantum Systems and ARC Centre of Excellence for Dark Matter Particle Physics, Department of Physics, University of Western Australia, 35 Stirling Highway, Crawley, WA 6009, Australia.}
\author{Gray Rybka}
\affiliation{Centre for Experimental Nuclear Physics and Astrophysics, University of Washington, 1410 NE Campus Parkway, Seattle, WA 98195, USA.}
\author{Michael E. Tobar}
\email{michael.tobar@uwa.edu.au}
\affiliation{ARC Centre of Excellence for Engineered Quantum Systems and ARC Centre of Excellence for Dark Matter Particle Physics, Department of Physics, University of Western Australia, 35 Stirling Highway, Crawley, WA 6009, Australia.}

\date{\today}

\begin{abstract}
Axions are a popular dark matter candidate which are often searched for in experiments known as ``haloscopes" which exploit a putative axion-photon coupling. These experiments typically rely on Transverse Magnetic (TM) modes in resonant cavities to capture and detect photons generated via axion conversion. We present a study of a resonant cavity design for application in haloscope searches, of particular use in the push to higher mass axion searches (above $\sim$60$\,\mu$eV). In particular, we take advantage of azimuthally varying TM$_{m10}$ modes which, whilst typically insensitive to axions due to field non-uniformity, can be made axion-sensitive (and frequency tunable) through strategic placement of dielectric wedges, becoming a type of resonator known as a Dielectric Boosted Axion Sensitivity (DBAS) resonator. Results from finite-element modelling are presented, and compared with a simple proof-of-concept experiment. The results show a significant increase in axion sensitivity for these DBAS resonators over their empty cavity counterparts, and high potential for application in high mass axion searches when benchmarked against simpler, more traditional designs relying on fundamental TM modes.
\end{abstract}

\maketitle

\section{\label{sec:level1} Introduction}
The composition and nature of dark matter continues to elude physicists, despite decades of observations implying its existence \cite{Zwicky2009, Rubin1980, ade2016planck}. 
However, the search for compelling candidates is narrowing through various experimental and theoretical efforts. In particular, the class of particles known as WISPs (weakly interacting sub-eV particles) are becoming increasingly favoured as dark-matter candidates \cite{Jaeckel2010}.
The axion is one such particle, widely considered amongst the most compelling dark-matter candidates, which arises as a consequence of an elegant solution to the strong \textit{CP} problem in QCD \cite{Peccei2008}. 

The proposal of the axion haloscope by Sikivie in 1983 was one of the first plausible methods of detecting axions in the laboratory by way of exploiting their expected coupling with photons \cite{Sikivie1983}. The inverse Primakoff effect is the mechanism by which an axion decays into a real photon through the absorption of another photon. If the cavity contains a geometrically appropriate resonant mode at the correct frequency, the signal will be resonantly enhanced. The power in the cavity can then be read out via a low-noise receiver chain. However, because the axion mass and the strength of its coupling to photons is unconstrained by theory, there exists a very large parameter space to be searched and many experiments are required to span the range. Several such experiments exist \cite{ADMX18, Brubaker2018}, with many focused around the microwave frequency band (corresponding to masses in the $\mu$eV range). However, many experiments are increasingly interested in lower-mass \cite{Ouellet2019} and higher-mass \cite{McAllister2017b}  axions.

The majority of the physics of the axion is determined by a parameter known as the Peccei-Quinn symmetry-breaking scale, $f_a$, which arises in the solution to the strong \textit{CP} problem that motivates the axions. $f_a$ is what most axion experiments ultimately hope to measure or constrain. This parameter is unconstrained by theory (although some cosmological constraints exist \cite{Burrows1990, Abbott1983}). $f_a$ determines the axion mass and the strength of its coupling to photons according to 

\begin{equation}
\begin{split}
\mathrm{m}_a&\sim\frac{4.51\times10^{15}}{f_a} \mathrm{eV},
\\
g_{a\gamma\gamma}&=\frac{g_\gamma\alpha}{f_a\pi}, 
\end{split}
\end{equation}
where $m_a$ is the mass of the axion, $g_{a\gamma\gamma}$ is the two-photon coupling constant of the axion and $\alpha$ is the fine-structure constant \cite{Kim1979, Kim2010, Dine1981}. The dimensionless axion-model-dependent parameter $g_{\gamma}$ is of order one and takes different values in different axion models. In the most popular two models, the Kim-Shifman-Vainshtein-Zakharov (KSVZ) and Dine-Fisher-Srednicki-Zhitnisky (DFSZ) models, $g_\gamma$ takes values of $-0.97$ and 0.36, respectively \cite{Kim1979, Kim2010, Dine1981}.

To date, the Axion Dark Matter eXperiment (ADMX) is the most sensitive and mature haloscope experiment, placing impressive exclusion limits on the searchable parameter space \cite{Asztalos2010, Hoskins2011, Braine2020}. However, current ADMX cavity designs are limited to probing masses of the order of a few $\mu$eV at KSVZ and DFSZ sensitivity. 

Currently, the high-axion-mass regime ($>60\, \mu$eV or $15\,$GHz) is largely inaccessible using traditional haloscope designs, this being attributed to the substantial decrease in sensitivity in this mass range owing to a range of technical factors that will be discussed below. Interestingly, despite the lack of sensitive experimental constraints, this high-mass region has benefited from a recent surge in theoretical and observational motivation \cite{Beck2013, Beck2015, Ballesteros2017, Berkowitz2015}. For example, the SMASH model favours axions with mass $\sim 100\,\mu$eV \cite{Ballesteros2017}, while other more recent models predict a much stronger axion-photon coupling \cite{co2020predictions, co2020leptoaxiogenesis, luzio2020landscape, Di_Luzio_2017, 1Di_Luzio_2017}. 

As discussed, haloscopes operate on the principle that axions from the galactic dark-matter halo are resonantly converted into detectable photons in a cavity. The signal power due to axion-photon conversion for a critically coupled cavity, with axion conversion occurring on resonance is given by \cite{Kim_2020}

\begin{equation}
\label{power}
P_a = g_{a\gamma\gamma}^2 \frac{\rho_a}{m_a} B_0^2 V C\, \mathrm{min}(Q_c,Q_a).
\end{equation}

The parameters $g_{a\gamma\gamma}$, $m_a$, local axion halo dark-matter density $\rho_a$ and the effective axion signal quality factor $Q_a \sim 10^6$, owing to the velocity distribution of dark matter are beyond experimental control. However, the external magnetic field strength $B_0$, the cavity volume $V$, the mode-dependent form factor $C$, and the cavity mode quality factor $Q_c$ are parameters within experimental control \cite{Kim_2020}. The form factor for a given mode in a cylindrical cavity, with a homogeneous static magnetic field aligned in the $\hat{z}$ direction can be defined as

\begin{equation}
\label{C}
C=\frac{\abs{\int dV_c \vec{E_c}\cdot \vec{\hat{z}}}^2}{V\int dV_c \epsilon_r \abs{E_c}^2},
\end{equation}

where $\vec{E_c}$ is the cavity electric field and $\epsilon_r$ is the relative dielectric constant of the medium. For a non-zero form factor, there must exist some degree of overlap between the electromagnetic field of the axion induced photon and electromagnetic field of the resonant cavity mode and the integral of this overlap must be non-zero. Thus, in an empty cylindrical cavity, only TM$_{0n0}$ modes will couple to axions in the experimental context outlined above. The highest form factor belongs to the TM$_{010}$ mode, which is consequently the mode of choice for most haloscope searches. 

$Q_c$ can be calculated through the mode-dependent geometry factor $G$ (assuming that resistive wall losses are the dominant factor, this being true for well-designed empty cavities and those utilizing low-loss dielectrics at cryogenic temperatures in normal-conducting materials).
\begin{align}
\label{g}
Q_c = \frac{G}{R_s} && G = \frac{\omega \mu_0 \int |\vec{H}|^2 dV}{\int |\vec{H}|^2 dS}
\end{align}

Here $R_s$ is the surface resistance of the material, $\vec{H}$ is the cavity magnetic field, $\omega$ is the resonant angular frequency of the cavity mode and $\mu_0$ is the vacuum permeability. It is assumed throughout this work that resistive wall losses are the dominant loss mechanism, far greater than any losses in low-loss dielectric materials, such as sapphire \cite{sapphireloss}. 

As mentioned, poor constraints on the axion mass and photon coupling strength create a large searchable parameter space. This places a high premium on axion-sensitive haloscopes with frequency tuning mechanisms. We therefore define the scanning rate of a haloscope as \cite{Kim_2020}

\begin{equation}
\label{scan}
\frac{df}{dt} = \frac{g_{a\gamma\gamma}^4}{\mathrm{SNR}^2} \frac{\rho_a^2}{m_a^2}\frac{B_0^4V^2C^2}{k_B^2 T_s^2}\frac{\beta^2}{(1+\beta)^2}Q_a\,\mathrm{min}(Q_L,Q_a),
\end{equation}

where $SNR$ denotes the chosen signal-to-noise ratio, $T_s$ represents the total system noise temperature, largely due to the noise of the first-stage amplifier and $Q_L=Q_c/(1+\beta)$ is the loaded quality factor, in which $\beta$ is the coupling strength of the receiver. The sensitivity of an experiment is therefore measured by the rate at which a haloscope can scan through a frequency range, at a desired level of axion-photon coupling and signal-to-noise ratio. The figure of merit for resonator design is then given by the quantity $C^2V^2Q_L$ or, equivalently $C^2V^2G$, as these are the controllable parameters which explicitly depend on the chosen resonator.

Now we can see why axion haloscopes become increasingly difficult at high masses. The volume, $V$ of resonant cavities scales by $V \propto f^{-3}$, and the expression contains an explicit dependence on ${m_a}^{-2}$. Furthermore, the noise temperature, $T_s$ of amplifiers increases at higher frequency, and the surface resistances of materials increase leading to a decrease in $Q_L$. All of these factors conspire to decrease $\frac{df}{dt}$ rapidly with increasing axion mass, making haloscope searches extremely difficult, requiring careful resonator design. Some suggestions on how to mitigate this problem at high frequencies include multiple cavity designs \cite{McAllister2017b, Brubaker2018}. 
\section{\label{sec:level2} Dielectric Haloscopes}

Dielectric embedded haloscopes, first proposed in Ref. 
\cite{Morris1984}, have been of growing interest in recent times. Since axion conversion has a high dependence on field geometry, the addition of dielectric in suitable regions can alter the geometry to favour axion conversion. Experiments such as MADMAX \cite{Caldwell2017}, the Electric Tiger \cite{Phillips2017}, and DALI \cite{DALI} incorporate dielectrics to facilitate their axion searches, for various reasons.

\begin{figure*}[htb]
\centering
    \begin{subfigure}[h]{0.46\textwidth}
    \centering
      \includegraphics[width=0.85\textwidth]{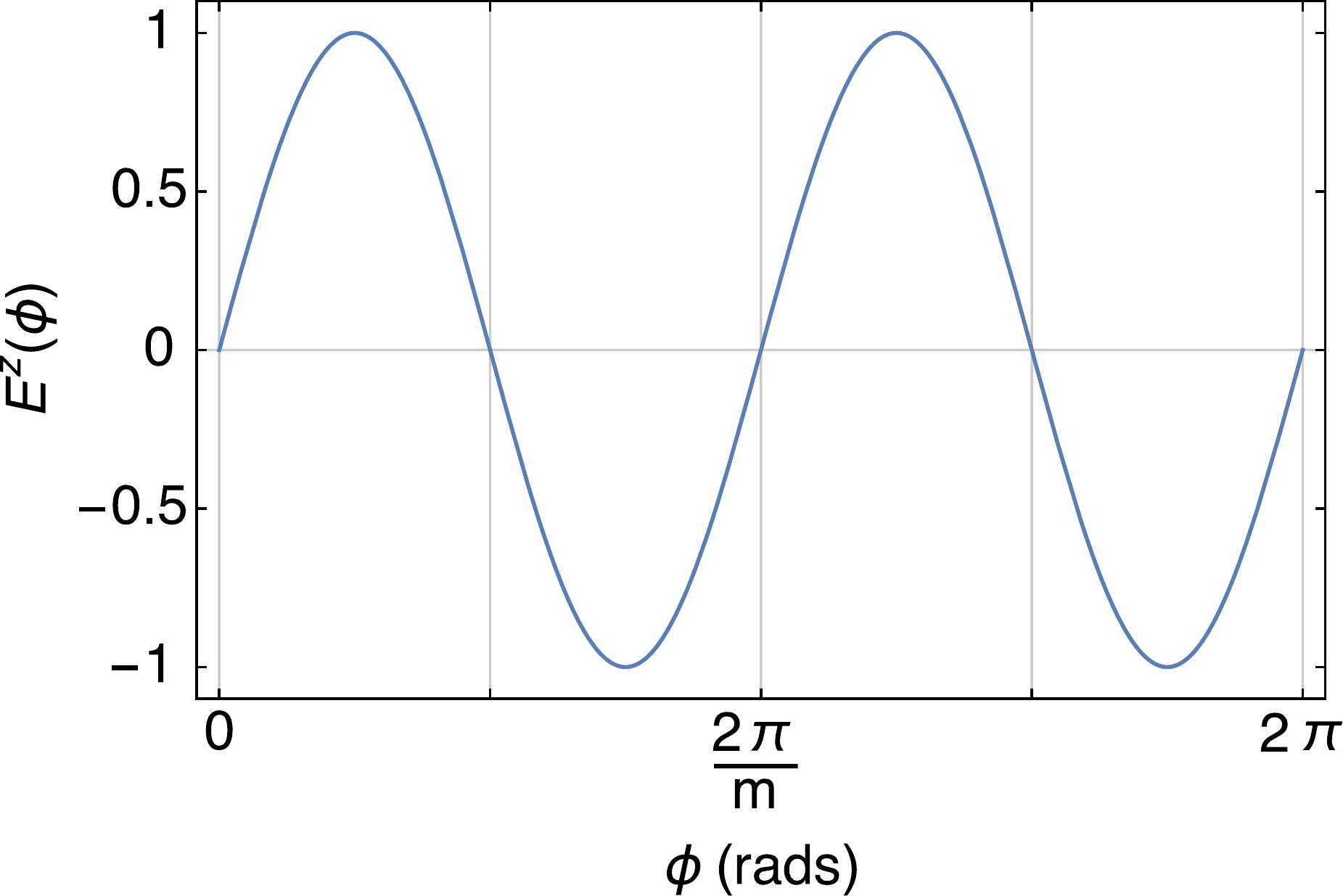} 
    \end{subfigure}      
\hsmash{\scalebox{2}{\stackon{$\rightarrow$}{\scriptsize\textsf{}}}}
    \begin{subfigure}[h]{0.49\textwidth}
    \centering  
      \includegraphics[width=0.85\textwidth]{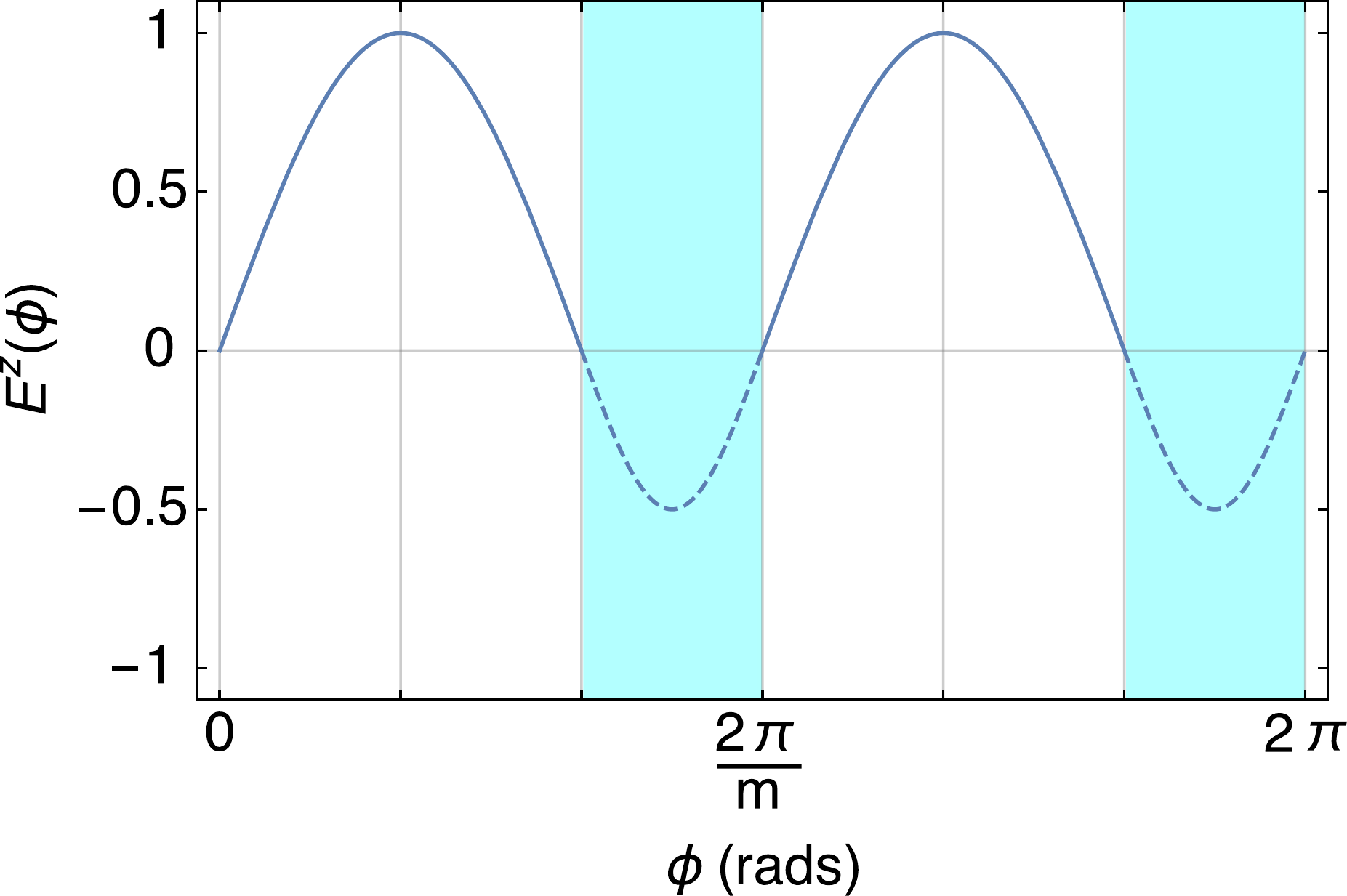}
    \end{subfigure}
     \caption{The $E_z(\phi)$ field of a TM$_{m10}$ mode shown before and after the addition of dielectric wedges (blue regions), as viewed in the azimuthal direction.\label{sketch}}
    \end{figure*}

Traditional, tuning rod cavity haloscope designs, such as the ones used by ADMX \cite{Asztalos2010, Hoskins2011}, exploit the TM$_{010}$ mode for its superior form factor of $\sim 0.69$. However, resonators that utilise lower order modes are ineffective at higher frequencies due to the dramatic decrease in volume, since the cavity dimensions must be of order $\lambda/2$, where $\lambda$ is the axion's Compton wavelength (which decreases with mass). Higher order resonances are thus attractive in the push to probe higher axion frequencies, allowing for higher cavity volumes at a given frequency. The cost of using higher order modes is the large degree of field variation, resulting in degraded form factors, cancelling out the sensitivity benefit from the increased volume. For example, compared to a TM$_{010}$ mode in an empty cavity, a TM$_{020}$ has a significant portion of its E$_z$ field out of phase with the applied $B$ field, reducing the coupling between the cavity mode and the axion, and degrading the form factor to $\sim 0.13$.  However, this issue can be addressed with the use of carefully placed dielectric materials to alter the field structure of the higher order modes, as has been previously considered for the case of TM$_{0n0}$ modes with $n>1$. 

As shown by McAllister et al. (2017) \cite{McAllister2017a}, careful placement of dielectric ``rings" in out of phase regions of the E$_z$ field of higher order TM$_{0n0}$ modes successfully mitigates this loss in form factor, while keeping the cavity volume high. This is possible due to the fact that dielectric structures effectively suppress electric field. Additionally, TM$_{010}$ modes are highly uniform, which makes frequency tuning difficult due to the high degree of symmetry. The use of dielectrics can be exploited to create ``built-in" tuning mechanisms as a result of more free parameters and broken symmetries in the cavity. Such resonators have been named Dielectric Boosted Axion Sensitivity (DBAS) resonators in the context of TM$_{0n0}$ modes. This has been further confirmed by some recent experiments by Kim et al. \cite{JKim20}, who introduced further ways to tune such TM$_{0n0}$ modes with reasonable frequency tunability. Also, Alesini et al. \cite{Alesini20}   have recently realised a fixed frequency prototype for axion searches, with boosted quality factor. In this work we consider another type of DBAS resonator, in the context of TM$_{m10}$ modes with $m>0$, and will refer to them as the Wedge DBAS resonators. This resonator appears visually similar to a dielectric equivalent to the multiple cell ``Pizza'' resonator proposed recently \cite{Jeong2018}. In this case, the wedges act like the boundaries of the individual ``Pizza''  cells.

\section{\label{sec:level3} Wedge DBAS Resonators}

The cavity mode electric field, $\vec{E_c}$  for a given TM$_{m10}$ mode inside a hollow cylindrical resonator of radius $R$, parametrised in cylindrical coordinates $r$, $\phi$ and $z$ is defined as 

\begin{equation}
\vec{E_c}= E_0\;e^{i\omega t}\;{J}_{m}\left(\frac{\varsigma_{m,1}}{R}{r}\right)\text{cos}({m\phi})\;\hat{z}.
\end{equation}

Where $E_0$ is some constant denoting the amplitude of the field and ${J}_{m}$ is a Bessel J function of order $m$, with  $\varsigma_{m,1}$ denoting its 1$^{\mathrm{st}}$ root (ie., the cavity wall). The field is in one phase in the $r$ direction, but alternates in phase $m$ times in the $\phi$ direction over the $2\pi$ range. Therefore, implementation of the DBAS method for a given TM$_{m10}$ mode would require the placement of $m$ dielectric wedges in the $m$ lobes of one of the phases, suppressing their contribution to the form factor integral shown in equation \ref{C} by suppressing the field amplitude in these regions.

Maximisation of the out of phase $E_z$ field confinement inside the dielectric wedges is done by placing the dielectric boundaries of the wedges between nodes of the field. For a TM$_{m10}$ mode, each of the $m$ total azimuthal variations occurs over a range of $\frac{2\pi}{m}$ rad; in this range, the field must alternate between maxima in both phases.
 We denote the optimal dielectric region size by $\theta$ (the angular size of each wedge) and the region without dielectric by $\bar{\theta}$ (vacuum). Hence we can find $\theta$ by demanding that
\begin{equation} 
\theta + \bar{\theta} = \frac{2\pi}{m}.
\label{eq:theta}
\end{equation}
The introduction of dielectric material reduces the speed of light within it by a factor of $\sqrt{\epsilon_r}$. This is tantamount to the space inside the dielectric increasing by a factor of $\sqrt{\epsilon_r}$, and so the physical size of the dielectric wedge must be decreased by this factor to meet our optimal condition. In the empty cavity structure, the angular size of the two phases is equal, and here we are reducing only one of them, such that $\theta = \bar{\theta}/\sqrt{\epsilon_r}$. Considering~\eqref{eq:theta}, the optimal dielectric wedge thickness, $\theta$, can then found to be
\begin{equation}
\theta = \frac{2\pi}{m(1+\sqrt{\epsilon_r})}.
\label{eq:wedgethick}
\end{equation}

\begin{figure*}[ht]
\centering
\centerline{
\includegraphics[width=0.7\textwidth]{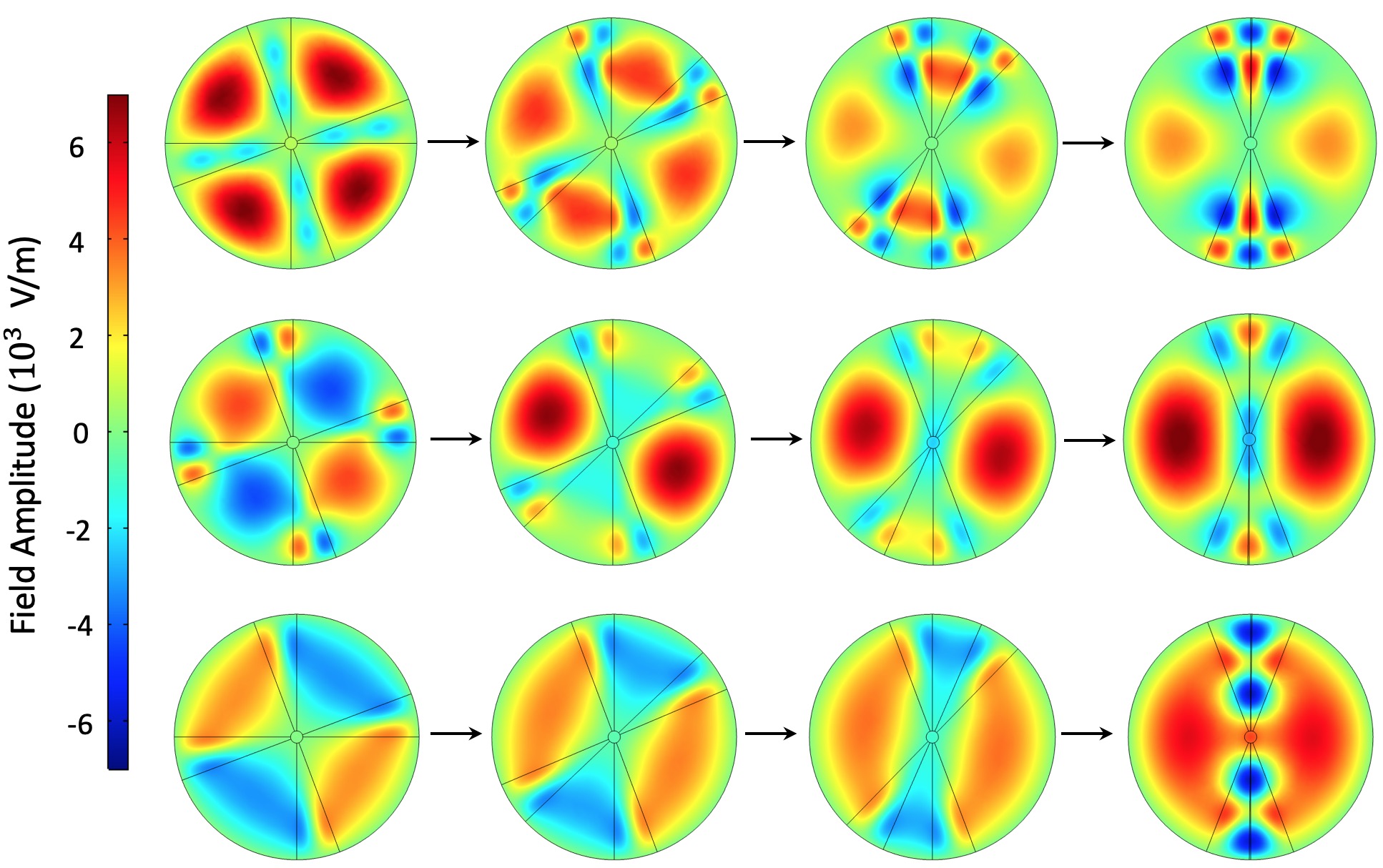}}
\caption{The $E_z$ profile of the TM$_{410}$ (upper), TM$_{410}$-like (middle) and TM$_{210}$ (lower) modes in a four-sapphire-wedge cavity as the wedges move together. It should be noted that the dielectric cylinder at the center acts only to increase the minimum mesh size in that area and has little to no impact on the field structure. Each mode is shown at tuning angles of $\phi=0, \;0.4,\; 0.8$ and 1.2 rad from left to right respectively. \label{4wedge}}
\end{figure*}

\begin{figure*}
\centering
    \begin{subfigure}[t]{0.48\textwidth}
      \includegraphics[width=0.9\textwidth]{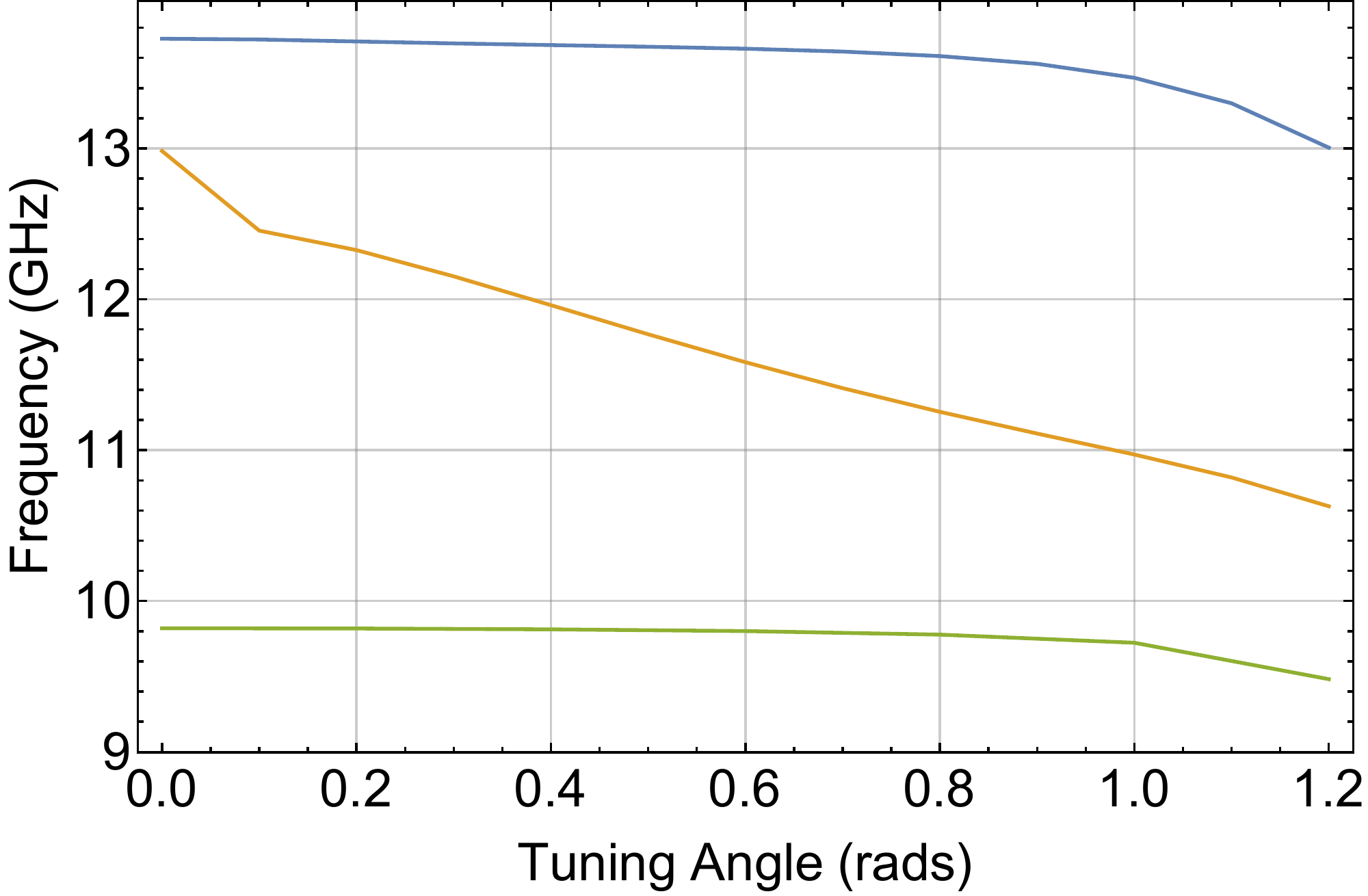} 
    \end{subfigure}
    \hfill
    \begin{subfigure}[t]{0.48\textwidth}
      \includegraphics[width=0.9\textwidth]{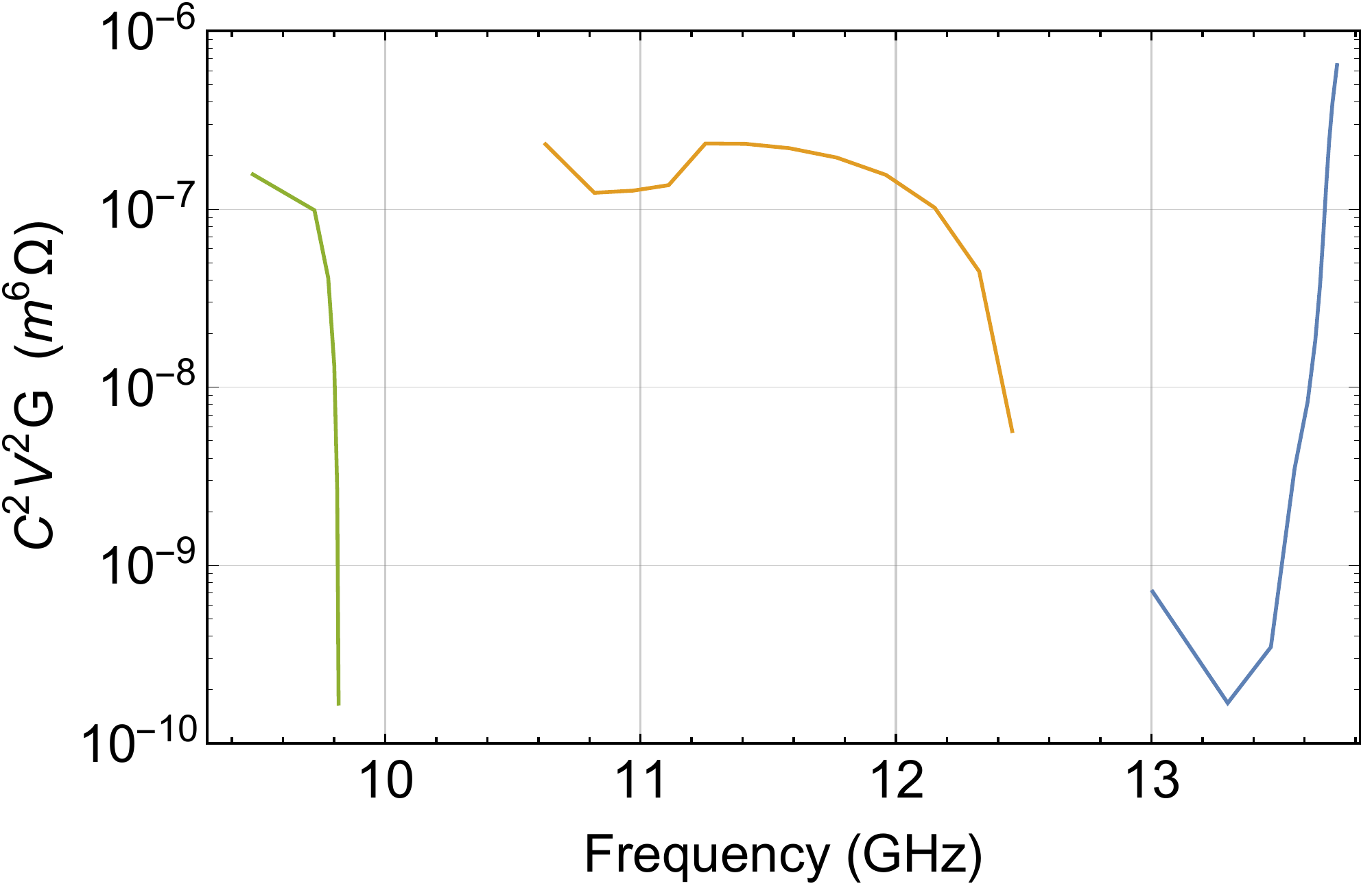}
    \end{subfigure}
     \caption{Left: the resonant frequencies of the TM$_{410}$ (blue), TM$_{410}$-like (orange) and TM$_{210}$ (green) modes shown as a function of tuning angle, $\phi$. Right: $C^2V^2G$ product as a function of frequency for the three modes of interest.\label{4wedgetune}}
    \end{figure*}

Figure \ref{sketch} shows the implementation of the Wedge DBAS method by placing dielectric (blue regions) of appropriate thickness in the out of phase parts of the $E_z$ field. It should be noted that this sketch is not to scale and only serves to show the effects of adding dielectric; namely the suppressed amplitude of the $E_z$ field and the reduced size of the dielectric region as compared to the empty region (vacuum). The integration of $E_z$ over the entire range now produces a non-zero value, and hence a non-zero form factor. 
\section{Modelling}
\subsection{Four-Wedge DBAS cavity}
Using Finite Element Method (FEM) modelling in COMSOL Multiphysics, we investigate the axion sensitive TM modes in a four-wedge resonator, with sapphire chosen as the dielectric, and wedge sizes as per~\eqref{eq:wedgethick} with $m=4$. Potential axion haloscope mode candidates must be highly tunable while retaining a sufficiently high scan rate, as indicated by the product $C^2V^2G$, computed via the FEM. The ``built-in" frequency tuning mechanism for the wedge-type cavities explored in this work relies on tuning via the relative angular separation of the wedges from one another. This can be achieved in an $m$-wedge cavity with $m/2$ wedges that are stationary, and $m/2$ wedges that can move relative to the others. We denote the tuning parameter (degree of wedge rotation) by $\phi$ and we define $\phi=0$ as the starting symmetric position where the angular separation between all wedges is the same. The maximum possible tuning is given by $\bar{\theta}$ rad, defined from $\phi=0$ to the angular position where the wedges are touching. We find $\bar{\theta}\sim 1.21$ using equation \ref{eq:wedgethick}, where for sapphire, $\epsilon_r\sim 11.349$. Various axion sensitive, tunable modes exist in the structure, and will now be named, discussed and compared in turn.
\subsubsection{TM$_{410}$ mode}
Shown in the top panel of fig. \ref{4wedge} is the field structure of the TM$_{410}$ mode as the wedges tune together from left to right. Although resemblant of a TM$_{410}$ mode, strictly speaking, this mode is not a true TM$_{410}$ due to the intruding dielectric. Using the results from FEM and equation \ref{C}, we determine the form factor for this mode at each $\phi$ position. We find that the TM$_{410}$ mode successfully confines out of phase lobes of the E$_z$ field to produce a non-zero form factor, $C\sim0.37$ at the $\phi=0$ position, decreasing to $C\sim 0.0055$ at the $\phi = 1.2$ position.
\subsubsection{TM$_{410}$-like mode}
In an empty cavity, there exist degenerate doublet TM$_{m10}$ modes that are a quarter period out of phase, or $\frac{\pi}{2m}$. However, when dielectric is added and azimuthal symmetry is broken, the modes break degeneracy and move to different frequencies. Through the FEM, we find the formerly degenerate doublet of the TM$_{410}$ mode, with a similar but distinctly different field structure. This mode is therefore referred to as the TM$_{410}$-like mode, the field structure of which is shown in the second panel of fig.  \ref{4wedge}. It should be noted from our discussion on form factor that it is only after perfect symmetry between the wedges breaks ($\phi \neq 0$), that this mode becomes axion sensitive. Past this initial position, the form factor gradually increases to $\sim 0.2$ at the $\phi = 1.2$ position.

\begin{figure*}[ht]
\centering
\centerline{
\includegraphics[width=0.7\textwidth]{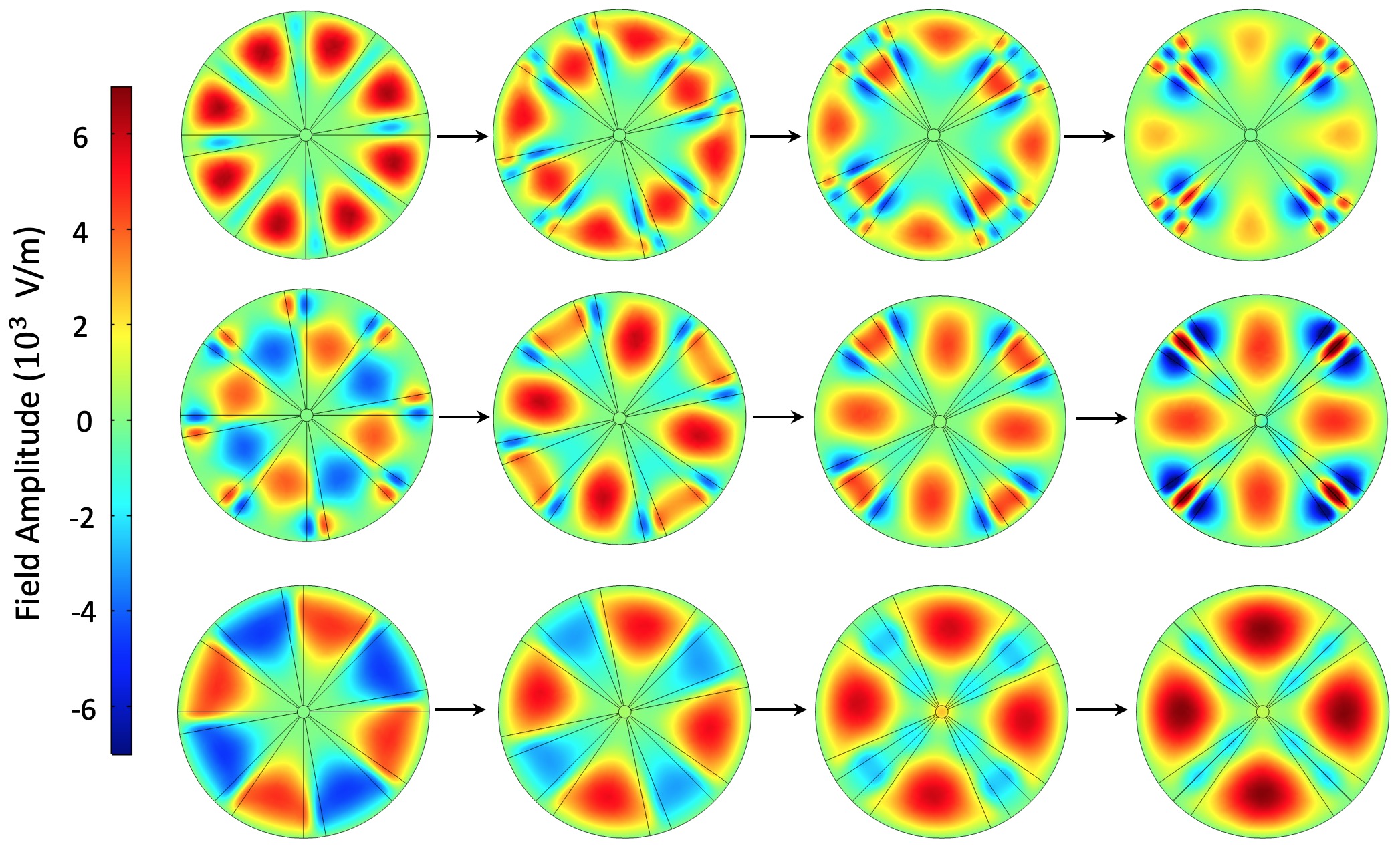}}
\caption{The $E_z$ profile of the TM$_{810}$ (upper), TM$_{810}$-like (middle), and TM$_{410}$ (lower) modes in a eight-sapphire-wedge cavity as the wedges move together. Each mode is shown at tuning angles of $\phi=0, \;0.2,\; 0.4$ and 0.6 rad from left to right respectively. \label{8wedge}}
\end{figure*}

\begin{figure*}[ht]
    \begin{subfigure}[t]{0.48\textwidth}
      \includegraphics[width=0.9\textwidth]{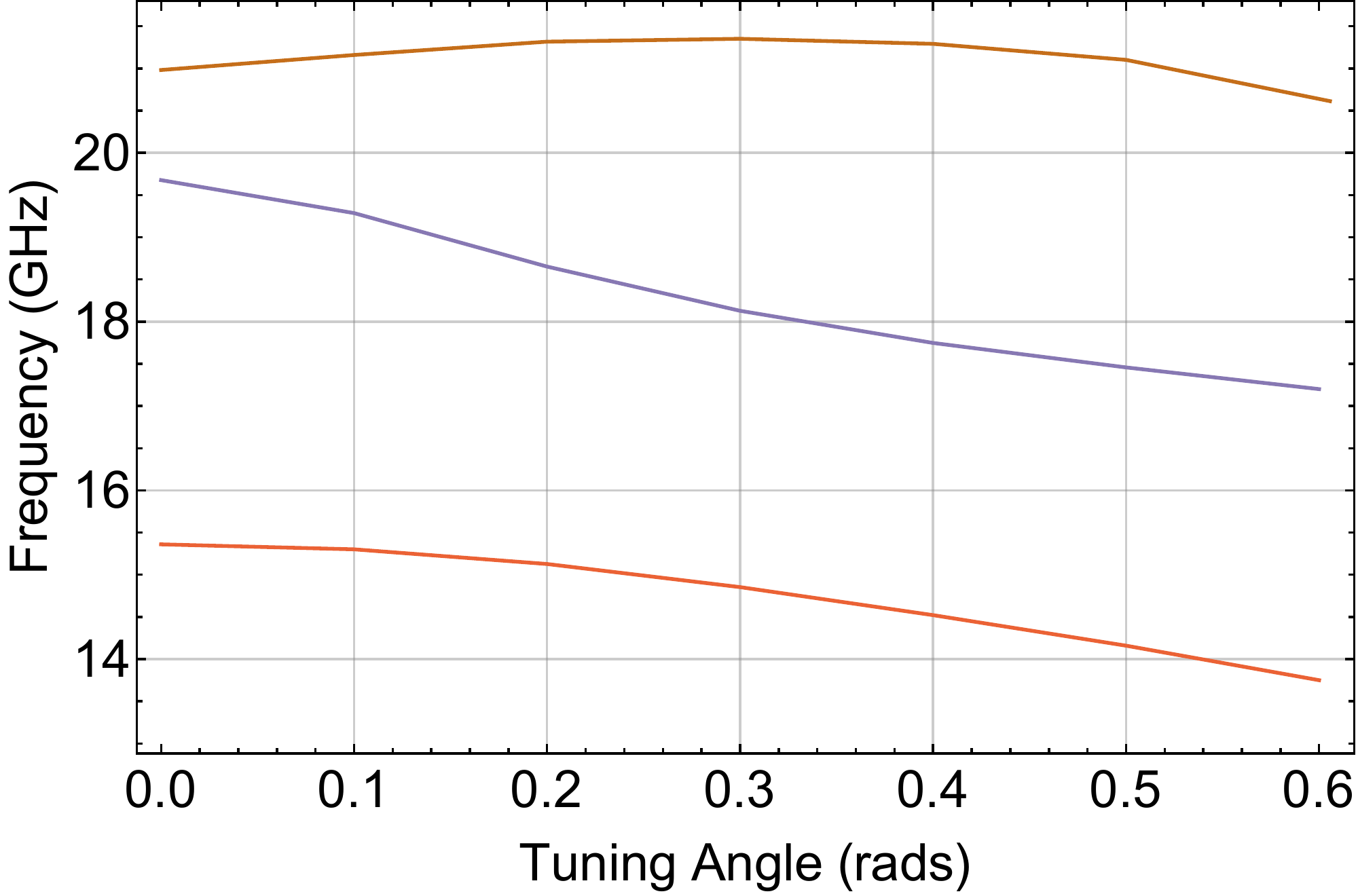} 
    \end{subfigure}
    \hfill
    \begin{subfigure}[t]{0.48\textwidth}
      \includegraphics[width=0.9\textwidth]{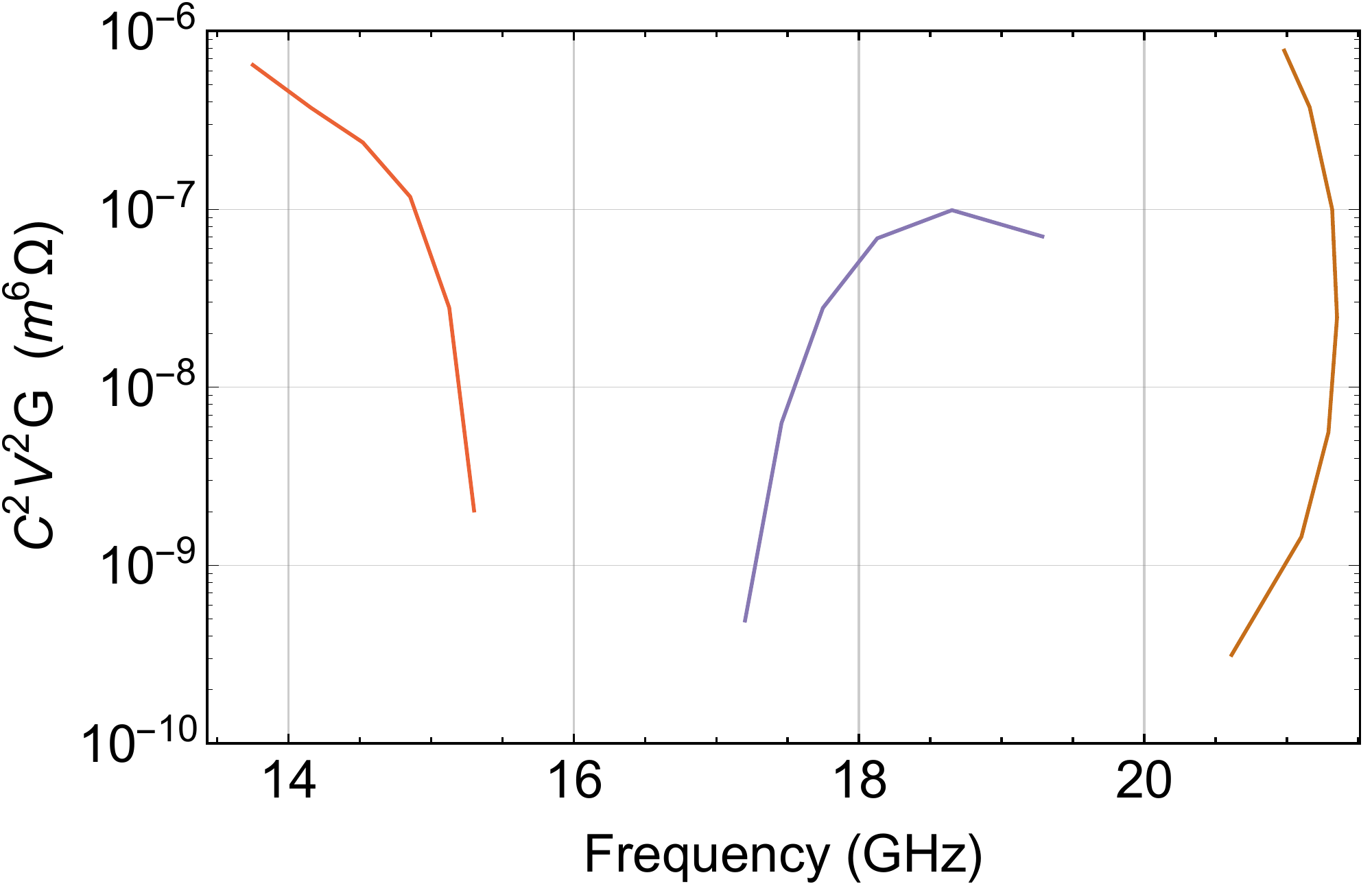}
    \end{subfigure}  \caption{Left: the resonant frequencies of the TM$_{810}$ (brown), TM$_{810}$-like (purple) and TM$_{410}$ (red) modes shown as a function of the tuning angle ,$\phi$ in an eight-sapphire-wedge cavity. Right: the $C^2V^2G$ product as a function of frequency for the three modes of interest.\label{8wedgetune}}
    \end{figure*}
    
\subsubsection{TM$_{210}$ mode}
DBAS $m$-wedge resonators that have $m\geq 4$ and even can once again exploit azimuthal symmetry to find TM$_{\frac{m}{2}10}$ modes, that tune in the same way as the previous two modes. These fractional modes only show significant sensitivity when the wedges are close together. This is an intuitive result, since we can think of an $m$-wedge resonator with its wedges tuned together as effectively being a $\frac{m}{2}$-wedge resonator, with an axion sensitive TM$_{\frac{m}{2}10}$ mode. Indeed, the optimal wedge angle for the TM$_{\frac{m}{2}10}$ mode is exactly double that of the TM$_{m10}$ mode. Through the FEM we find the field structure of the TM$_{210}$ mode in a four-wedge cavity as shown in the bottom panel of fig. \ref{4wedge}. It is clear that this mode begins with a form factor of zero and gradually becomes more sensitive as tuning progresses, increasing to have $C\sim 0.13$ at the $\phi=1.2$ position. 

\subsubsection{Sensitivity}    
The relevant axion sensitivity and frequency tuning results of FEM modelling for a four-wedge DBAS cavity with a radius of $20\,$mm, a height of $60\,$mm, and an angular wedge thickness ($\theta$) of $\sim 0.36$ rads are shown in fig. \ref{4wedgetune}. Although the total tuning range is shown, it is useful to define a so called ``sensitive" tuning range, which only considers a given mode when it is within an order of magnitude of the maximum $C^2V^2G$. The TM$_{410}$ mode shows the greatest peak sensitivity; however, tuning of this mode is poor, with a starting frequency of $\sim 13.7\,$GHz, we observe a total and sensitive tuning of $\sim$720 and $\sim 70\,$MHz respectively. The ``doublet" TM$_{410}$-like mode, however is a much more promising candidate for axion searches, offering substantial sensitivity over a broad tuning range, with a total and sensitive tuning of $\sim 2.4$ and $\sim1.7\,$GHz respectively. Analogous to the TM$_{410}$ mode, the TM$_{210}$, although sensitive to axion detection, also suffers from a poor degree of tuning, reporting only a total and sensitive tuning of $\sim$340 and $\sim320\,$MHz respectively.

\subsection{Eight-Wedge DBAS cavity}
\begin{figure}[htb]
\centering
\centerline{
\includegraphics[width=0.45\textwidth]{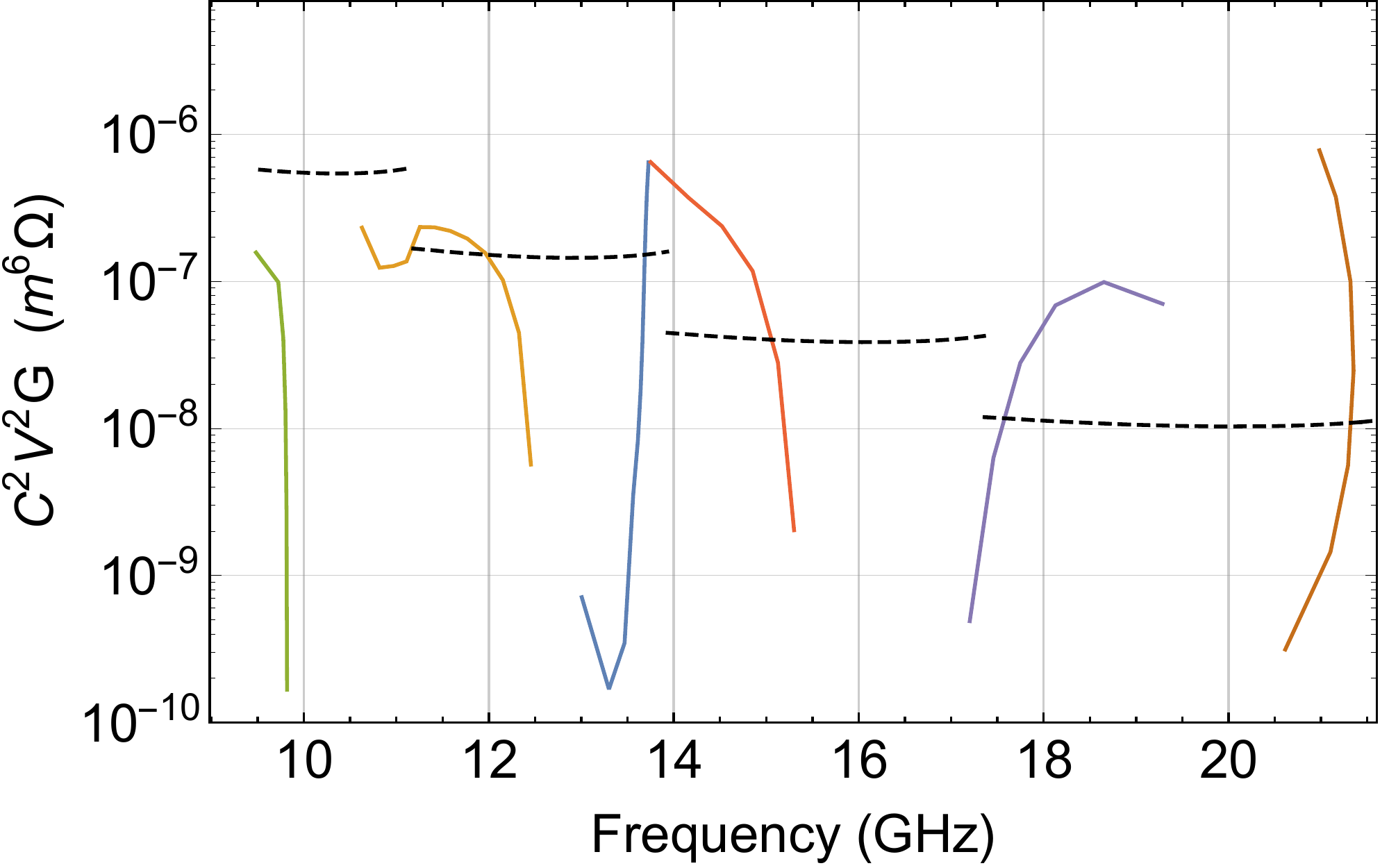}}
\caption{The prodcut $C^2V^2G$ as a function of frequency for the TM$_{810}$ (brown), TM$_{810}$-like (purple), TM$_{410}$ (red), TM$_{410}$ (blue), TM$_{410}$-like (orange) and TM$_{210}$ (green) modes. Shown in black (dashed) are a series of traditional conducting rod-tuned resonators, centred in different frequency regions for comparison.\label{allwedgesens}}
\end{figure}

Successfully finding three axion-sensitive modes in the four-wedge configuration, we now investigate higher order TM modes in an eight-wedge configuration, while using the same cavity dimensions. The angular size of each wedge will then be exactly half the size used in the four-wedge resonator, as shown in Eq.\ref{eq:wedgethick}. The results of the FEM modelling in an eight-wedge DBAS cavity are shown in fig. \ref{8wedge} and \ref{8wedgetune}. While not surprising, it is clear from the field profiles, that the same three modes of interest (with twice the number of azimuthal variations) also exist in the eight-wedge cavity, where the fractional TM$_{\frac{m}{2}10}$ mode is now a TM$_{410}$, as shown in the bottom panel of fig. \ref{8wedge}. We can then think of the TM$_{410}$ mode as having two tuning regimes; one in the earlier presented four-wedge cavity and another in the eight-wedge configuration. Importantly, this mode is axion-sensitive across different regions of frequency space for the two tuning regimes, effectively extending the mode's sensitive tuning range. Again, the fractional TM$_{410}$ mode starts with a form factor of zero, increasing to have $C\sim 0.37$ at the $\phi=1.2$ position, equivalent to the $\phi=0$ position in the four-wedge cavity. Interestingly, the  TM$_{410}$ mode performs better in the eight-wedge configuration when it comes to the total ($\sim1.6\,$GHz) and sensitive ($\sim1.3\,$GHz) tuning.

In similarity to the modes presented in the previous 4-wedge iteration, the TM$_{810}$ has $C\sim 0.33$ to be  maximal at the starting $\phi=0$ position, whereas the TM$_{810}$-like mode is completely axion-insensitive at this position. As tuning progresses, the TM$_{810}$ becomes less sensitive with $C$ decreasing to $\sim 0.003$, while the   TM$_{810}$-like mode increases to have a maximum $C\sim 0.11$. The total and sensitive tuning for the TM$_{810}$ mode is poor, with $\sim$740 and $340\,$MHz respectively. In contrast the TM$_{810}$-like mode has a more impressive tuning of $\sim$2.5 and $\sim1.8\,$GHz respectively.

\subsection{Practicalities and Mode Crossings}
As previously discussed, novel cavity design is an essential step in the push towards searching the higher frequency axion parameter space. However, there is an inherent trade off between cavity volume and the use of higher order modes. The DBAS method seeks to rectify this by mitigating the downside of higher order modes (reduction in form factor), while keeping the cavity volume high. However, higher order modes should be approached with caution, as they introduce significant mode crowding and risk ``avoided level crossings", resulting in degraded axion-sensitivity in those regions of frequency space. Mode crowding can be mitigated by reducing the cavity length, thus increasing the frequencies of higher order length dependent ($p\neq0$) modes with respect to the length independent modes. The cavity length can be optimised through extensive FEM modelling such that mode crossings are reduced to some acceptable level in the region of interest. Given this is not a design study for a specific frequency band, this type of aspect ratio optimisation was beyond the scope of this work. As a result, we opt for a relatively low and realistic aspect ratio of 3. For typical haloscopes, an aspect ratio of $\sim 5$ is common. 

\section{Possible Implementation And Comparison}
In principle, it would be possible to combine the eight and four wedge cavities discussed here. If we begin with the eight-wedge configuration, and tuned the wedges until they are touching, each pair of two wedges will be the same size as the wedges in the four-wedge cavity. We could then tune two of these new, thicker wedges relative to the other two, and recreate the tuning of the four-wedge cavity. In this way, all six axion-sensitive modes would become accessible within a single cavity. Since FEM modelling for both configurations is done using the same cavity dimensions, we plot $C^2V^2G$ against frequency for all six modes, as shown in fig. \ref{allwedgesens}. Although possible in principle, a modular Wedge DBAS design is highly conceptual and would face significant practical challenges in its implementation, foremost of which is an intricate tuning mechanism such that the eight-wedge configuration can ``fold" into four wedges, and then be tunable afterward. Alternatively, one could avoid significant engineering and complexity by simply inserting the desired wedge configuration, since the cavity radius is the same for both regimes. 

\begin{figure}[htb]
\centering
\centerline{
\includegraphics[width=0.45\textwidth]{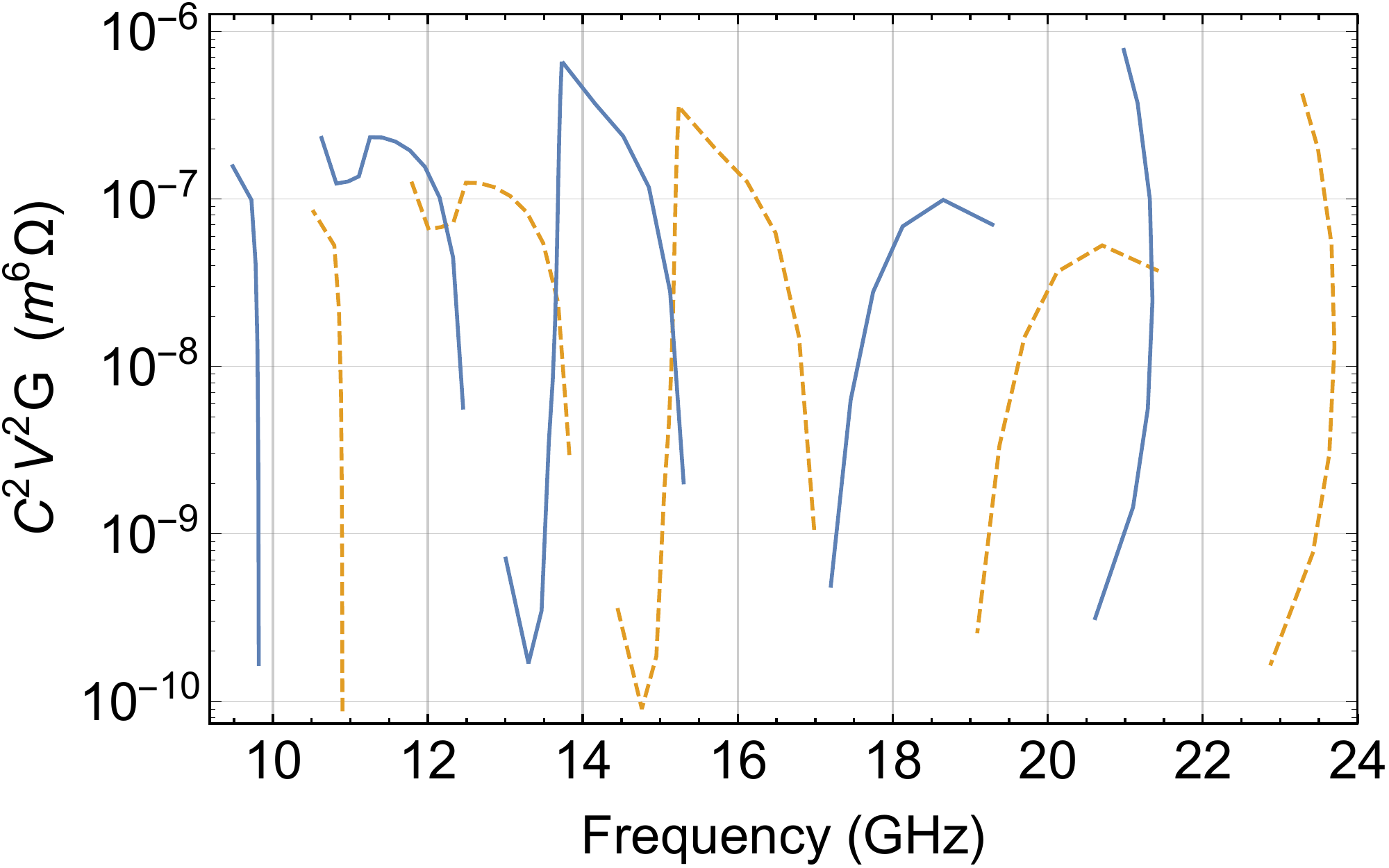}}
\caption{The product $C^2V^2G$ as a function of frequency for the relevant modes of interest with a scaled second cavity (dashed orange) presented containing the same modes at higher frequencies. \label{2ndcav}} 
\end{figure} 

To assess the viability of different haloscope designs, it is common practice to compare against a reference cavity that tunes in the same frequency range. We choose to benchmark against a TM$_{010}$ mode tuned by radially moving a conducting rod, resulting in subtle changes to the mode geometry, thus altering the resonant frequency. This is the type of resonator used by world-class haloscopes, and thus it is a good comparison for other designs. Overlaid in black (dashed) in fig. \ref{allwedgesens} is the $C^2V^2G$ data for this benchmark cavity, constructed and additionally scaled such that the frequency tuning ranges are comparable with the other wedge cavity designs presented. Consequently, the four regions of black dashed lines correspond to four tuning rod cavities with slightly different dimensions. To create a clear comparison, an aspect ratio of 3 was also chosen for the benchmark cavity design. There are ultimately many free parameters, and much optimisation possible in the design of both schemes, and thus the compared designs should be thought of as a relatively simple one. However it should be noted that the benchmark cavity, although comparable in sensitivity in these regions, becomes increasingly impractical to implement in the high frequency regime, attributed to the significantly reduced cavity and tuning rod dimensions.

\begin{figure*}[htb]
\centering
\centerline{
\includegraphics[width=0.7\textwidth]{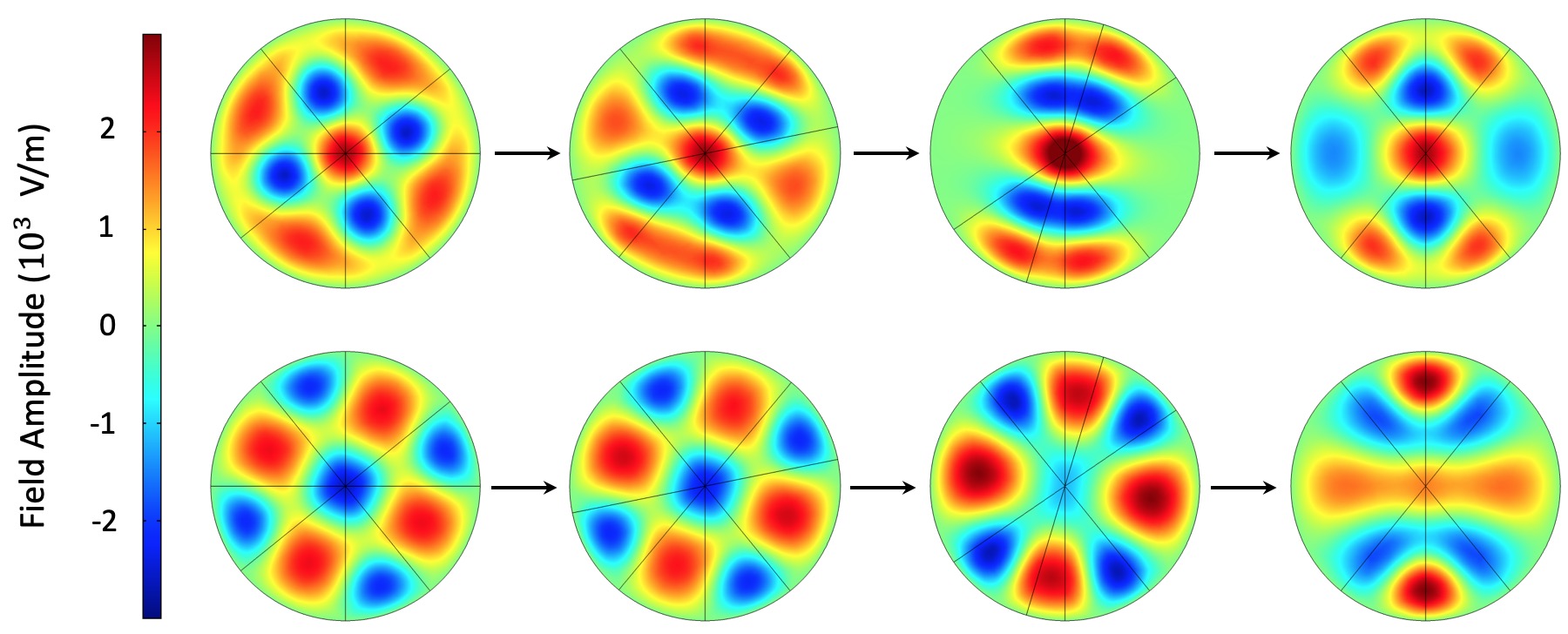}}
\caption{The $E_z$ profile of the TM$_{410}$-like (upper) and TM$_{410}$ (lower) modes in a four-Teflon-wedge cavity as the wedges move together. Each mode is shown at tuning angles of $\phi=0, \;0.2,\; 0.6$ and $\sim$0.93 rad from left to right respectively.\label{tefboth}}
\end{figure*}

As shown, almost all of the modes of interest have a peak $C^2V^2G$ greater than the benchmark cavity design, with some modes sustaining an improved scan rate over their entire sensitive tuning range. If implemented, this modular cavity design is very attractive as an axion haloscope in the hard to reach but well motivated high mass regime, due to its broadband tuning and high sensitivity, but we can take it even further.

Using the inverse relationship between radius and frequency ($\omega \propto R^{-1}$), we can simply scale the results from the modelled cavity to imitate the results of a second cavity with a slightly different radius, so that the gaps in the previous sensitivity plot (fig. \ref{allwedgesens}) are filled by a second resonator of the same type. As a result, two cavities of slightly different radii can be used to almost completely cover a frequency range between 9.5 and$21.5\,$GHz with a high degree of sensitivity. In the case of uniform rescaling, the volume changes with the cube of the radius, whereas mode dependent factors $C$ and $G$ remain unchanged. Therefore increasing the resonant frequency of a particular mode by a factor $f$ results in $C^2V^2G$ decreasing by a factor of $f^6$. The second cavity is scaled such that the resonant frequencies for the modes of interest increase by a factor $f=1.11$, degrading $C^2V^2G$ by $f^{-6}\sim0.53$. Once again it is clear why many experiments have so far been unable to probe the higher frequency parameter space. 

As shown in fig. \ref{2ndcav}, the combination of two multi-stage cavities (or effectively four, if the multi-stage design cannot be implemented and instead two sets of wedges are required at each radius) can almost completely cover a $12\,$GHz region with $C^2V^2G$ greater than $10^{-8}$. However, to get a more quantitative representation of the relative performance of each cavity design over a given frequency range, we compute the relative scan time $\Delta\tau$, which corresponds to the time taken for a haloscope to exclude a given frequency range at a given $g_{a\gamma\gamma}$ and SNR. The scan time is proportional to the following expression,  

\begin{equation}
\Delta\tau \propto \int_{f_1}^{f_2}\frac{1}{C^2V^2G(f)}\mathrm{d}f.
\end{equation}

The comparison for both cavity designs is done over the frequency range between 9.5 {}\textendash{} $21.5\,$GHz, and the relative scan time is normalised to the benchmark cavity, assuming all other parameters are equal. However, while the benchmark cavities span the entire frequency range, the Wedge cavities fall marginally short, and so the relative scan time should be normalised to the total searchable frequency range. With this taken into account, we find that the Wedge DBAS cavities outperform the benchmark cavities by a factor of $\approx$ 2.1.

\section{Proof of Concept Experiment}
To assess the viability of a Wedge DBAS type resonator, a prototype with four Teflon wedges is first considered. As a proof of concept this cavity is expected to tune the TM$_{410}$ and doublet TM$_{410}$-like modes in line with expectations from the COMSOL modelling (within experimental uncertainty). Teflon wedges are an ideal choice due to their relatively low cost and ease of production, unlike more expensive, harder to machine, low-loss crystals such as sapphire. Based on the success of the Teflon proof of concept, a sapphire resonator will be constructed and tested. 

\begin{figure}[htb]
\centering
\centerline{
\includegraphics[width=0.23\textwidth]{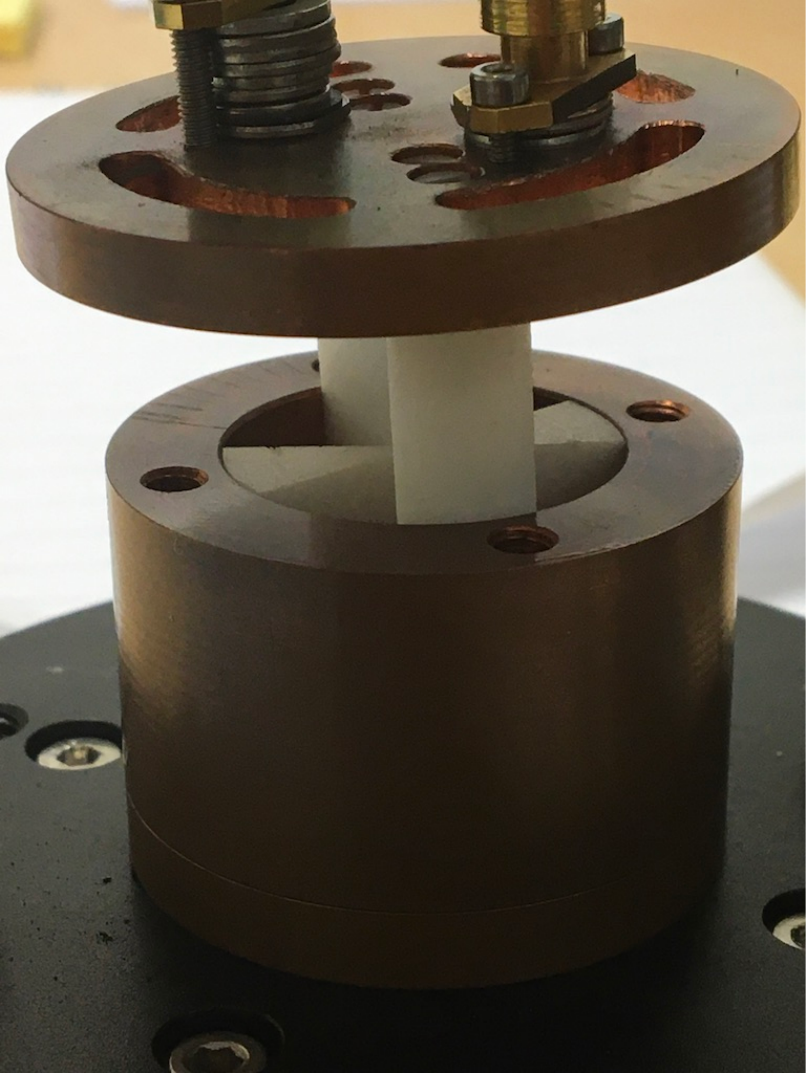}}
\caption{The Teflon Wedge DBAS cavity used in the proof of concept experiment. As discussed in the text, a pair of diametrically opposed wedges are mounted to the moveable lid, while another pair are affixed to the base.\label{cavitypic}}
\end{figure}

We use equation \ref{eq:wedgethick} to once again find the optimal Teflon-wedge thickness. The copper cavity has a radius of $13.47\, \mathrm{mm}$ and a height of $22.5\, \mathrm{mm}$. The field profiles for the modes of interest in the Teflon cavity are shown in fig. \ref{tefboth} and closely resemble what is seen in the sapphire iteration, albeit with a significant reduction in the degree of out of phase field suppression and the presence of a central lobe, attributed to Teflon's comparatively low permittivity, $\epsilon_r\sim2.1$. The TM$_{210}$ mode is not shown here and is not investigated further due to an absence of tuning, as indicated by the initial FEM results. 

\begin{figure}[htb]
\centering
\centerline{
\includegraphics[width=0.5\textwidth]{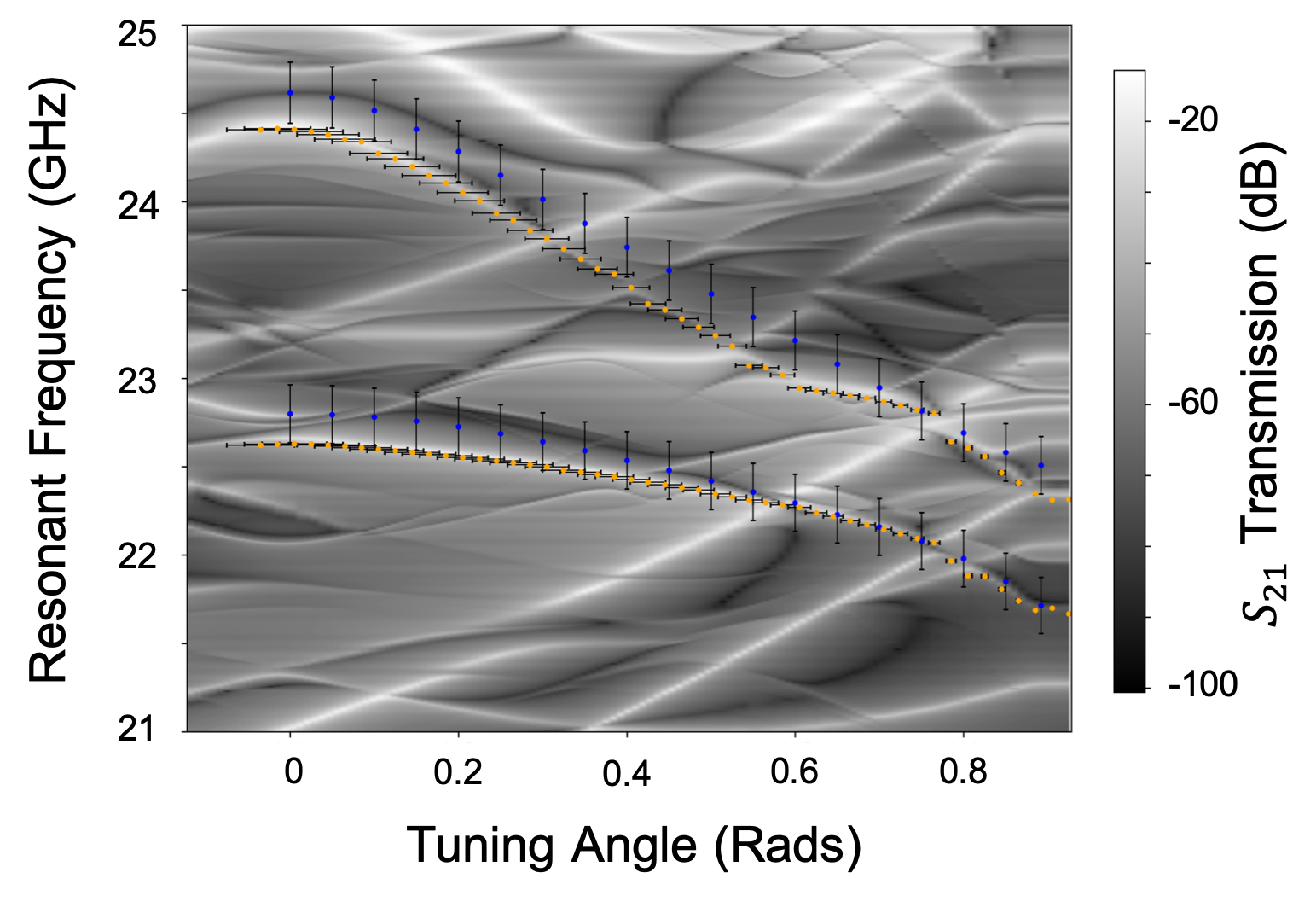}}
\caption{A colour-density plot of the transmission coefficient as a function of resonant frequency and the tuning angle $\phi$. Of specific interest are the TM$_{410}$ (lower) and $TM_{410}$-like modes (upper), identified by hand-taken measurements (orange) and the predicted tuning from FEM (blue). Darker regions represent less transmission while lighter regions represent greater transmission. Error bars represent uncertainty in the measured and predicted values, and are discussed in the text. \label{cdplot}}
\end{figure}

The proof of concept measurements are done at room temperature using a Thorlabs stepper motor and rotation stage. The lid of the cavity is clamped down, such that the base of the cavity is able to rotate relative to the lid. Two of the wedges are affixed to the base, and two to the lid, meaning that as the base tunes with respect to the lid, two of the wedges tuned. The cavity is coupled to coaxial antennae, and transmission measurements are made with a Vector Network Analyzer as a function of the wedge's angular position. The two modes are first tracked by hand using a step size of 0.02 rad, and later via automated transmission coefficient measurements that use a 0.0087 rad step size. The modelled and measured frequencies are shown as a function of the tuning angle in fig. \ref{cdplot}. The horizontal error bars placed on the measured data are due to the Thorlabs rotation stage quoting an accuracy in angular position of $\pm 820\, \mu$rad. This being an open-loop system, the horizontal error compounds for each subsequent measurement. Additionally, deviations from perfect symmetry can significantly perturb the mode field structure and hence frequency. Unequal wedge sizes, wedge tilt, crude measurements of their thickness (within $\pm 0.02$ rad) and the addition of probes greatly effect the resonant frequency of the mode. The modelling of these small perturbations in COMSOL in conjunction with other uncertainties results in a total uncertainty of approximately $\pm150\,$MHz in the modelled frequencies ($<1\%$ of the central starting frequency). The vertical error bars on the FEM data represent this uncertainty.

The loaded Q-factors, measured at various wedge positions are found to be $\mathcal{O}(1000-2000)$. For this proof of concept experiment, given that it is at room temperature and Teflon is relatively lossy, the dielectric losses due to the wedges cannot be neglected as they can be for our cryogenic sapphire simulations. We thus follow a slightly different procedure for determining the expected Q-factors in the proof-of-concept experiment. Using the FEM, it is possible to find the electric energy filling factor ($P_e$) inside the Teflon for the two modes of interest, and the geometry factor for the conducting surfaces. The Q-factor for the resonator is then given by \cite{spherebragg},

\begin{equation}
Q^{-1} = P_e Tan\delta+\frac{R_s}{G}.
\end{equation}

Using the geometry factor, $\frac{G}{R_s}$ is calculated to be $\mathcal{O}(30\, 000$ -- $40\,000)$. If we take $Tan\delta=1.6 \cdot 10^{-4}$ as in \cite{spherebragg}, and use $P_e \approx 0.45$ (from simulations), we find the expected Q-factor for both modes over the tuning range to be $\mathcal{O}(9000$ -- $10000)$. However, it should be noted that optimisation of the Q-factors is not a primary goal for this proof of concept, resulting in little effort to improve them. Standard Q-improving techniques, such as barrel knife edges, silver plating and RF chokes can be employed in future cryogenic iterations of this experiment, using less lossy sapphire as the dielectric, to ensure that the dominant loss factor is surface losses. 

Importantly, the overall shape of the two modelled modes matches almost perfectly what is seen experimentally. We also observe highly responsive frequency tuning as a result of the ``built-in" tuning mechanism. Furthermore, the modes in the proof of concept cavity are at even higher frequencies than the modelled sapphire cavity, owing to the diameter of the available Teflon stock\textemdash and very few avoided level crossings are observed over the experimental tuning range. These factors demonstrate the viability of this promising resonator design. 
\\
\par
\section{Conclusion}
This work presents a theoretical and experimental study of a  Wedge DBAS cavity resonator for use in high mass axion haloscopes. Through strategic placement of dielectric structures, these resonator designs are shown to significantly boost the form factors of various TM$_{m10}$ modes. The FEM modelling results for both eight and four-wedge cavity configurations are presented, and show six axion-sensitive modes with varying levels of frequency tuning. We compare their performance with that of a conventional conducting rod resonator, and find the DBAS cavity modes to boast superior $C^2V^2G$ products, albeit each over a reduced tuning range. However, the viability of this resonator design is enhanced when both the eight and four-wedge regimes are combined into a single two-stage resonator, effectively broadening its sensitive tuning range by allowing access to all six modes within a single cavity. These modes are especially promising for applications at higher frequencies than those accessible with traditional rod-tuned haloscopes.

Also undertaken is a proof of concept experiment using a prototype Teflon-wedge DBAS cavity. The cavity’s “built-in” tuning mechanism is successful in altering the frequency for the modes of interest in a highly responsive and reliable way, demonstrating the feasibility of such a design. Currently, plans are in place to commission a cryogenic-compatible four-wedge cavity using less lossy, higher permittivity sapphire for possible implementation in the ORGAN Experiment. 
\\
\par
This was funded by the ARC Centre for Excellence for Engineered Quantum Systems, CE170100009, and the ARC Centre for Excellence for Dark Matter particle Physics, CE200100008, as well as ARC grant number DP190100071

\end{document}